\documentclass[jpcm]{iopart}
\usepackage{epsf,graphicx,citesort}
\usepackage[dvips]{color}
\definecolor{dred} {rgb} {0,0,0}
\newcommand{\dred}[1]{{\color{dred} #1}}
\newcommand{\gl}[1]{Eq. (\ref{#1})}
\newcommand{\gls}[2]{Eqs. (\ref{#1},\ref{#2})}
\newcommand{\glto}[2]{Eqs. (\ref{#1}) to (\ref{#2})}
\def\gtrless{\raise2.5pt\hbox{$>$}\llap{\lower2.5pt\hbox{$<$}}}
\def\gtrapprox{\raise2.5pt\hbox{$>$}\llap{\lower2.5pt\hbox{$\approx$}}}
\newcommand{\bsq}[1]{\begin{subequations}\label{#1}}
\newcommand{\esq}{\end{subequations}}
\newcommand{\beq}[1]{\begin{equation}\label{#1}}
\newcommand{\eeq}{\end{equation}}
\newcommand{\beqa}[1]{\begin{eqnarray}\label{#1}}
\newcommand{\eeqa}{\end{eqnarray}}
\newcommand{\wer}{\;\mbox{where}\quad}

\newcommand{\gd}{\dot{\gamma}}
\newcommand{\kapa}{\mbox{\boldmath $\kappa$}}
\newcommand{\kapl}{\mbox{$\cdot${\boldmath $\kappa$}}}
\newcommand{\kap}{\mbox{$\cdot${\boldmath $\kappa$}$\cdot$}}
\newcommand{\kapr}{\mbox{{\boldmath $\kappa$}$\cdot$}}
\newcommand{\kapt}{\mbox{$\cdot${\boldmath $\kappa$}$^T\cdot$}}
\newcommand{\kaprt}{\mbox{{\boldmath $\kappa$}$^T\cdot$}}
\newcommand{\smop}{\Omega}
\newcommand{\smopb}{\Omega^{\dagger}}

\newcommand{\vek}[1]{{\bf #1}}
\newcommand{\pa}{\mbox{\boldmath $\partial$}}
\renewcommand{\rho}{\varrho}

\begin{document}

\title{\dred{Integration through transients for} Brownian particles under steady shear}\footnote{This paper is dedicated to Professor Lothar Sch{\"a}fer
on the occasion of his 60th birthday.}

\author{M. Fuchs$^{1}$ and M. E. Cates$^2$}
\address{$^1$ Fachbereich Physik, Universit\"at Konstanz, D-78457
  Konstanz, Germany}  
\address{$^2$ School of Physics, JCMB Kings Buildings, The University of
Edinburgh, Mayfield Road, Edinburgh EH9 3JZ, United Kingdom}

\begin{abstract}
Starting from the microscopic Smoluchowski equation for interacting Brownian particles under
stationary shearing, exact expressions for shear--dependent steady--state averages, correlation and structure functions, and susceptibilities are obtained, which take the form of generalized Green--Kubo relations. They require integration of transient dynamics. Equations of motion with memory effects for transient density fluctuation functions are derived from the same microscopic starting point. We argue that the derived formal expressions provide useful starting points for approximations in order to describe the stationary non--equilibrium state of steadily sheared dense colloidal dispersions. 
\end{abstract}

\maketitle

\section{Introduction}

Colloidal dispersions can be driven into stationary non--equilibrium states by shearing. Their properties are important for the handling of dense colloidal dispersions, but yet not well understood from fundamental starting points \cite{russel}. A prominent and universally observed effect is {\em shear thinning}, that the viscosity of the solution decreases by orders of magnitude upon increasing the shear rate. A widely used many--body model of colloidal particles under shear is provided by the Smoluchowski equation \cite{dhont}, a special form of a Fokker--Planck equation \cite{Risken}, which, however, has yielded to exact solution only at low particle densities  \cite{Bergenholtz02}. There it exhibits weak shear thinning followed by shear thickening. The model supposes the existence of a given solvent velocity flow profile that depends linearly on distance along one direction, as has been observed in polydisperse dense fluid and glassy colloidal systems \cite{Derks04,Petekidis02b}, and thus it does not include changes of the solvent velocity field induced by the particle motion.   
\dred{(Nor does it allow for `shear banding' or other symmetry breaking phenomena.)}
An additional simplification of the model can be performed when solvent induced interactions (`hydrodynamic interactions') are neglected, so that the model effectively describes {\em interacting Brownian particles in a constant shear flow}.  Gratifyingly, shear thinning has been observed in simulations of this system \cite{Strating99}, where also homogeneous states in a linear flow profile were recorded. The model thus contains flow curves (viz. curves of stress versus shear rate) in qualitative agreement with typical experimental data of systems close to glassy arrest \cite{Senff99} at not too high shear rates (where hydrodynamic interactions presumably dominate \cite{russel}).

Recently we presented a mode--coupling approach which leads to a consistent and (in principle) parameter--free, quantitative, albeit approximate description of the stationary sheared state at high particle concentrations or strong interactions   \cite{Fuchs02}. It explains the behaviour of dense dispersions under shear from considering the competition of local caging of particles \cite{Goetze91b}, which causes slow structural relaxations, with shear advection of fluctuations \cite{Cates89}, which speeds up the decay. As an important concept it uses integration of the transient dynamics in order to gain insights into the stationary state \dred{presumed to be} reached at late times. A number of \dred{rather} universal predictions of the approach have already been obtained \cite{Fuchs03}, and are in qualitative agreement \cite{Fuchs03b} with, e.g., the mentioned computer simulations \cite{Strating99}. In this contribution we present details of the approach, starting from the many--body Smoluchowski equation with shear and setting up the frame for integrating through the transient. Formally exact expressions for stationary averages, correlation functions and susceptibilities, and for transient density correlators are presented. 
\dred{These exact results lay the foundations for our
approach to sheared colloids, whose approximations, outlined in Ref. \cite{Fuchs02}, will be given in detail in a companion paper. In particular the results derived below create a framework within which to make mode-coupling-type approximations for sheared colloids, without invoking the equilibrium form of the fluctuation-dissipation theorem (which cannot reliably be used under shear). 

It is well known in many physical situations that the same standard approximation (e.g., factorising an average) gives different results when applied to two formulations of a problem that would, if treated exactly, be equivalent. A careful choice is then required, and the work reported here can be thought of as `preparing the best ground' for a judicious mode-coupling approach to sheared colloids.}
 
The derived generalized Green--Kubo relations and generalized Zwanzig--Mori equations, which may \dred{also} be of interest on their own, are useful to describe the non--equilibrium steady state because they enable one to connect the stationary distribution function to the transient dynamics evaluated with equilibrium averaging.  This strategy was followed in mode--coupling calculations of the non--linear viscosity of simple liquids under shear, where the slow relaxation of `long time tails' leads to non--analytic dependences on shear--rate  \cite{Kawasaki73b}. Transient fluctuation functions were also successfully used in some of the simulation studies of this problem, and the connection to the theoretical  approach was shown explicitly \cite{Morriss87}. Here shearing cuts off the anomalous long--time dynamics present in the quiescent fluids, and subtle (but small) corrections to the viscosity arise. 

Recent mode coupling theory approaches to simple liquids close to glassy arrest by Miyazaki and Reichman
\cite{Miyazaki02,Miyazaki04}, and to violations of the fluctuation--dissipation theorem in Brownian particle systems by Szamel \cite{Szamel04}, follow a somewhat different approach \dred{from ours}. There, time--dependent correlation functions for fluctuations around the sheared steady state are obtained, as are susceptibilities describing the response of the state in Ref. \cite{Szamel04}.  Thus, in the spirit of the mode coupling theory of quiescent systems \cite{Goetze91b}, structural quantities of the stationary state, which now depend on shear rate, enter as input into the equations describing the dynamics. Importantly, Miyazaki and Reichman find that shear advection of density fluctuations speeds up the structural relaxation, which would become excessively slow close to glassy arrest. 

Our approach, as sketched in Ref. \cite{Fuchs02}, uses generalized Green--Kubo relations to access the stationary distribution function, in order to allow for its (possibly) non--analytic dependence on shear rate. We expect \dred{nonanalyticities} to arise when we consider the rheological properties of the system close to solidification into a colloidal glass, because the quiescent dynamics becomes non--ergodic at a glass transition described by the idealized mode coupling theory \cite{Goetze91b}. Therefore in this contribution, $(i)$ in Sect. {\it \ref{green}} stationary averages are reformulated so that the transient dynamics enters; in the companion paper \dred{we will present in detail} mode coupling approximations so that the transient dynamics is described by transient density fluctuation functions. Then $(ii)$ in Sect. {\it \ref{tradens}}, the equations of motion of transient density functions are reformulated in such a way that dense systems can be described, where particle interactions lead to large memory--effects. \dred{In Sect. {\it \ref{model}} the model is defined and some properties discussed, while Sect. {\it \ref{disc}} concludes with a short outlook to the companion paper.}

\section{Steady state properties}
\label{model}

\subsection{Microscopic starting point}

The system considered consists of $N$ spherical
particles (diameter $d$) 
dispersed in a volume $V$ of solvent with imposed flow
profile $\vek{v}(\vek{r}) = \kapr\,  \vek{r}$,  where for 
simple shear with velocity along the $x$-axis and
its gradient along the $y$-axis, the shear rate tensor is
$\kapa =\gd\  \hat{\bf x} \hat{\bf y}$
(viz. $\kappa_{\alpha\beta}=\gd \delta_{\alpha x}\delta_{\beta y}$).  The effect of the shear
rate $\gd$ on the particle dynamics is measured by the Peclet number
\cite{russel}, Pe$_0=\gd d^2/D_0$, formed with the (bare) diffusion
coefficient $D_0$ of a single particle.  
Dimensionless units are obtained by setting
$d=D_0=k_BT=1$, whereupon Pe$_0 = \gd$.
The evolution of the distribution function $\Psi(\Gamma)$
 of the particle positions, 
$\vek{r}_i$, $i=1,\ldots, N$ (abbreviated into
$\Gamma=\{{\bf r}_i\}$), under internal forces $\vek{F}_i =
- \pa_i U(\Gamma)$ 
(with the total interaction potential $U$) and
shearing, but neglecting hydrodynamic interactions, is given by the
Smoluchowski equation \cite{russel,dhont}: 
\beqa{a1}
\partial_t \Psi(\Gamma,t) & = & \smop(\Gamma) \; \Psi(\Gamma,t)\; ,
 \nonumber\\
\smop & = & \Omega_{e} + \delta \Omega = \sum_i \pa_i \cdot 
\left( \pa_i - \vek{F}_i - \kapr\, \vek{r}_i \right)  \; ;
\eeqa
here $\Omega_{e}= \sum_i \pa_i \cdot 
\left( \pa_i - \vek{F}_i \right)  $ abbreviates the Smoluchowski Operator (SO) without
shear. In the following, operators act on everything to the right, if not marked differently by bracketing.
The conditional probability, for the system to evolve from
state point $\Gamma'$ at time $t'$ to $\Gamma$ at the later time $t$,
denoted by $P(\Gamma t | \Gamma' t')$, also is determined from $\smop$:
\beq{a2}
\partial_t P(\Gamma t |  \Gamma' t')  =  \smop(\Gamma) \; P(\Gamma t |
 \Gamma' t')\; ,
\eeq
with the initial condition $P(\Gamma t | \Gamma' t) =
\delta(\Gamma-\Gamma')$, that both state points coincide at the
same time.

There exist two special time--independent distribution functions; 
the equilibrium one, $\Psi_e$, and the stationary one, $\Psi_s$, which satisfy:
\beq{a3}
 \Omega_e \Psi_e = 
 0 \qquad , \qquad \smop \Psi_s = 0 \; .
\eeq
The equilibrium one is determined from the total internal 
interaction energy $U$ via
the Boltzmann weight, $\Psi_e(\Gamma) \propto e^{- U(\Gamma)}$, 
but the stationary distribution function $\Psi_{s}$
 is unknown. Averages with $\Psi_e$ will be abbreviated by $ \langle
\ldots \rangle = \int d\Gamma \, \Psi_e(\Gamma) \ldots$, while  
$\Psi_s$ determines steady state averages, denoted by
 $\langle \ldots\rangle^{(\gd)}= \int d\Gamma \, \Psi_s(\Gamma) \ldots$.

At finite shearing, steady--state averages $f$, time--dependent correlation $C_{fg}(t)$
and time--independent structure
functions $S_{fg}$, for fluctuations $\delta f = f - \langle
f \rangle^{(\gd)}$ around the steady state,  and 
response susceptibilities $\chi_{fg}(t)$, are the central objects of
interest:
\beqa{a4}
f(\gd) &=& \langle f \rangle^{(\gd)}  = \int d\Gamma\; \Psi_{s}(\Gamma)\; f(\Gamma)   \nonumber\\
C_{fg}(t) &=& 
 \int d\Gamma\int d\Gamma'\; W_2(\Gamma t+t',\Gamma' t') \; \delta f^*(\Gamma')\; \delta g(\Gamma)
\nonumber\\ &=&
\langle \delta f^* \; e^{\smopb t}\; \delta g \rangle^{(\gd)} 
  \nonumber\\
S_{fg} &=& C_{fg}(t=0) = \langle \delta f^* \; \delta g \rangle^{(\gd)} 
  \nonumber\\
\chi_{fg}(t) &=& \langle\,  \sum_i \; \frac{\partial f^*}{\partial {\bf r}_i} \cdot \pa_i \;
e^{\smopb t}\; g \rangle^{(\gd)}  \; .
\eeqa
The calculation of the fluctuation functions involves the joint probability distribution,
$W_{2}(\Gamma t, \Gamma' t')$, 
that the system is at point $(\Gamma,t)$ after  it was in a stationary state at
$(\Gamma',t')$;
it is given using the conditional probability that is the solution of \gl{a2}: 
\[W_2(\Gamma t,\Gamma' t') = P(\Gamma t | \Gamma' t') \; \Psi_{s}(\Gamma')= e^{ \smop(\Gamma) (t-t') } \; \delta(\Gamma-\Gamma') \;\Psi_{s}(\Gamma')\; .\] 
Exchanging the order of times, $t'>t$, it obeys $W_2(\Gamma t,\Gamma' t') = 
W_2(\Gamma' t',\Gamma t)$.
The adjoint of the SO arose in the fluctuation function $C(t)$ 
from partial integrations:
\beq{a5}
\smopb = \sum_i ( \pa_i + \vek{F}_i + \vek{r}_i \kapl^T)  \cdot  \pa_i\; ,
\eeq
where surface contributions are neglected, throughout, for the considered infinite system ($V\to\infty$).
The susceptibility $\chi_{fg}(t)$ describes the linear change of the expectation value of variable $g$:
\beq{aa5}
\Delta g(\gd)(t) = \langle g \rangle^{(\gd , h_e)} - \langle g \rangle^{(\gd)}
= \int_{-\infty}^t\!\!\!dt' \; \chi_{fg}(t-t') \; h_e(t') + {\cal O}(h_e^2) \; ,
\eeq
upon application of an external field $h_e(t)$ that couples to the variable $f^*$ in the potential energy; viz. when the potential energy $U$ is perturbed to:
 \beq{aa6}
U \to U - f^*(\Gamma)\; h_e(t) \; .
\eeq
The standard calculations \cite{Risken} leading from \gl{aa6} to \gl{a4} are sketched in the Appendix.

Without applied shear the SO $\smopb_e$ is an Hermitian operator with respect to equilibrium averaging \cite{dhont} 
\beq{a55}
\langle (\smopb_e f^*) g \rangle =\langle f^* \smopb_e g \rangle = - \sum_{i}\; \langle  \frac{\partial f^*}{\partial \vek{r}_i} \cdot
 \frac{\partial g}{\partial \vek{r}_i} \rangle\; ,
\eeq
and (as seen from specialising to $f=g$)  possesses a negative semi--definite spectrum.
But with shear $\smopb$ cannot be brought into an Hermitian form \cite{Risken}; see Sect. {\it \ref{eigen}} below. 
The action of $\smop$ on the equilibrium distribution function $\smop \Psi_e=\delta \smop \Psi_e$
 will become important later on and allows one to define the stress tensor:
\beqa{a6}
\delta \smop\, 
 \Psi_e &=& 
- \sum_i \pa_i \; \kap \; {\bf r}_i \; \Psi_e 
=
 - \sum_i \left( \vek{F}_i \kap \,\vek{r}_i + {\rm Trace}\!\{\kapa\} \right) \;
\Psi_e \nonumber \\ &=&  {\rm Trace}\!\left\{\kapa \; \mbox{\boldmath $\sigma$} \right\}\;  \Psi_e = \gd \ \sigma_{xy}\; \Psi_e ,
\eeqa
with $\sigma_{\alpha\beta}$ the zero--wavevector  limit of the 
potential part of the stress tensor:
\beq{a7}
\sigma_{\alpha\beta} = - 
\sum_i \; ( \delta_{\alpha\beta} + F^\alpha_i r^\beta_i ) 
\; .
\eeq 
The specific form of $\kapa$ for sheared systems was used in the last equality of \gl{a6} only. 

\subsection{Basic properties}

Some well known properties of solutions of
Fokker--Planck equations \cite{Risken} shall be collected which bear relevance to the
discussion of sheared colloidal dispersions. 

\subsubsection{Eigenfunctions expansions: }
\label{eigen}

The Smoluchowksi equation of \gl{a1} may be viewed as a continuity equation in phase space \cite{Risken},
$\partial_t \Psi(\Gamma,t) + \sum_i \pa_i \cdot {\bf J}_i(\Gamma,t) = 0$, where the probability current equals:
\beq{b1}
{\bf J}_i(\Gamma,t)  = \left( {\bf F}_i + \kapr {\bf r}_i - \pa_i \right) \Psi(\Gamma,t) \; .
\eeq
Stationarity implies $\sum_i \pa_i \cdot {\bf J}_i(\Gamma,t\to\infty) = 0$. But only if the current vanishes, ${\bf J}_i =0$, in the steady state,  
one can show that the SO $\smop$ is related to an Hermitian operator. 
Necessary condition for ${\bf J}_i =0$ are the 'potential conditions':
\beq{b3}
\partial_i^\alpha \left( { F}_j^\beta + (\kapr {\bf r}_j)^\beta \right) = \partial_j^\beta \left( {F}_i^\alpha + (\kapr {\bf r}_i)^\alpha  \right) 
\eeq
While the potential conditions hold in equilibrium, they are violated under shear, because
$\partial (\hat{\bf y}\cdot\kapr {\bf r}_i))/\partial x_i \ne
\partial (\hat{\bf x}\cdot\kapr {\bf r}_i))/\partial y_i $. 
\dred{Then}, if an expansion in eigenfunctions 
of the SO exists, it will have the following properties: The conditional probability from \gl{a2} takes the form 
\beq{b4}
P(\Gamma t | \Gamma' t') = \sum_{n} \; \varphi_n(\Gamma) \; \hat{\varphi}_n(\Gamma') \; e^{- \lambda_n \, (t-t') }\; .
\eeq
where the eigenvalues satisfy $\Re{\lambda_n} \ge0$, and the sets of eigenfunctions 
\beq{b5}
\smop \; \varphi_n = - \lambda_n \; \varphi_n \; \quad\mbox{ and }\; 
\smopb \; \hat{\varphi}_n = - \lambda^* \; \hat{\varphi}_n \; ,
\eeq
are bi--orthogonal, viz.: 
\beq{b6}
\int d\Gamma \hat{\varphi}_n(\Gamma) \; \varphi_m(\Gamma) = \delta_{n  m} \; .
\eeq
Yet, because no further connection between the sets of eigenfunctions exists in general, important properties of 
equilibrium fluctuations cannot be expected under shear: for example the autocorrelation functions  $C_{ff}(t)$ of \gl{a4}  can  fail to be of positive type and may exhibit negative frequency spectra \cite{McLennan88}.

\subsubsection{Fluctuation Dissipation Theorem: }

In the case without shear, where $\smop_e$ is the SO and averages are performed with the equilibrium distribution function $\Psi_e$, a simple relation exists between the fluctuation function $C_{fg}^{(e)}(t)$ and the susceptibility $\chi_{fg}^{(e)}(t)$.
By partial integration, and recalling that $\pa_i \Psi_e = {\bf F}_i \Psi_e$, one finds:
\beqa{b7}
\chi_{fg}^{(e)}(t) &=&  - \langle \delta f^* \; \smopb_e \; e^{\smopb_e \, t}\; \delta g \rangle \nonumber \\
&=&  - \, \partial_t \; C_{fg}^{(e)}(t)\; .
\eeqa
The expected fluctuation dissipation theorem (FDT) connects response and fluctuation function. On the other hand, with   shear 
the susceptibility $\chi_{fg}(t)$ is connected to a fluctuation function of a variable $\tilde f$, which can only be found if the stationary distribution function is known \cite{Risken,Szamel04}; it satisfies:
\beq{b8}
\sum_i \; \pa_i \cdot \frac{\partial f^*}{\partial {\bf r}_i} \Psi_s = \smop\;  \tilde{f}^* \;
\Psi_s\; ,
\eeq
and the FDT then states \cite{Risken}:
\beq{b9}
\chi_{fg}(t) =  - \langle \delta \tilde{f}^* \; \smopb \; e^{\smopb\, t}\; \delta g \rangle^{(\gd)} =
-  \partial_t \; C_{\tilde{f}g}(t)\; .
\eeq
This appears not to be as useful as \gl{b7}.
 
\subsubsection{Aspects of translational invariance: }
\label{tra}

Homogeneous amorphous systems shall be studied so that by assumption the equilibrium distribution function  $\Psi_e$ is translationally invariant and isotropic. As shown in Sect. {\it \ref{trans}},
 the steady--state distribution function with shear, $\Psi_s$, then also \dred{is translationally invariant, assuming that no   spontaneous symmetry breaking takes place, but anisotropic.} 
Appreciable simplifications follow for steady--state quantities of
 wavevector--dependent fluctuations:
\beq{b11}
f_{\bf q}(\Gamma,t) = e^{\smopb \, t} \; \sum_i\; X^f_i(\Gamma) \; e^{i\vek{q}\cdot {\bf r}_i}\; ,
\eeq
where e.g. $X^\rho_i=1$ describes density fluctuations $\rho_{\bf q}(t)$, while $X^\sigma_i
=\delta_{\alpha\beta} + (1/2) \sum_j' (r^\alpha_i-r^\alpha_j) du(|{\bf r}_i-{\bf r}_j|)/dr^\beta_i$ 
gives the stress tensor element $\sigma_{\alpha\beta}({\bf q})$ for interactions described by the pair--potential $u$.
Translational invariance in an infinite sheared system dictates that averages
are independent of identical shifts of all particle positions, $\Gamma \to 
\Gamma'$ where $\vek{r}_i' = \vek{r}_i + \vek{a}$ for all $i$. Under such a shift
the SO becomes
\beq{b115}
\smopb(\Gamma) = \smopb(\Gamma') - {\bf a} \; \kapt \; {\bf P } \; , \quad \mbox{ with }\, 
{\bf P} = \sum_i \pa_i \; .
\eeq
Thus a fluctuation of a variable which depends on particle separations only, 
 viz.  $X^f_i(\Gamma) = X^f_i(\Gamma')$ so that ${\bf P} X^f_i(\Gamma)=0$ 
holds, transforms to
\beq{b12}
f_\vek{q}(\Gamma,t) = e^{- i \left( \vek{q} + {\bf q} \kapl \; t \right)\cdot \vek{a} }
\; f_\vek{q}(\Gamma',t) \; .
\eeq

As the integral over phase space must agree for either integration variables $\Gamma$  or $\Gamma'$, 
 steady--state averages from \gl{a4} can be non--vanishing for zero wavevector only: 
\beq{b13}
f_{0}(\gd)\;  \delta_{\bf q , 0} =
\frac 1V\; \langle  f_{\vek{q}}(t) \rangle^{(\gd)} \; .
\eeq 
In the following the index $0$ will often be suppressed in e.g. the average density $\rho=N/V$ and the shear
stress $\sigma(\gd) = \langle \sigma_{xy}\rangle^{(\gd)}/V$ from \gl{a7}. Also, because mostly finite wavevectors
will  be considered, the nonzero averages at $q=0$ often will be suppressed so that we have for  fluctuations `$\delta f_{\bf q}= f_{\bf q}$'.
 
Similarly, wavevector--dependent steady--state structure functions from \gl{a4} obey:
$S_{f_\vek{k} g_{\bf q}} = N S_{fg; {\bf q}} \delta_{\bf k , q}$, where
\beq{b14}
S_{fg; {\bf q}}(\gd) = \frac 1N \; \langle  \delta f^*_\vek{q} \; \delta g_\vek{q}  \rangle^{(\gd)}\; .
\eeq
The familiar structure factor built with density fluctuations shall be abbreviated by
$S_{\bf q}(\gd)= \frac 1N \; \langle  \rho^*_\vek{q} \; \rho_\vek{q}  \rangle^{(\gd)}$. 
While these findings are familiar from systems without shear, translational invariance of sheared systems
takes a special form for the two--time correlation functions from \gl{a4}. Because 
\[
C_{f_\vek{k} g_{\bf q}}(t) = e^{- i \left( \vek{q} \kapl\, t + \vek{q} - \vek{k} \right)\cdot
 \vek{a} }\; C_{f_\vek{k} g_{\bf q}}(t)  \;,
\]
as follows from \gl{b12}, a fluctuation with wavevector $\bf q$ is correlated with a fluctuation
of ${\bf k}={\bf q}(t)$ with the {\em advected} wavevector ${\bf q}(t) = {\bf q} + \vek{q} \kapl\, t$ at the later time $t$; only then the exponential in the last equation becomes unity. The advected wavevector's
 $y$--component increases with time as $q_y(t)=q_y+\gd\, t\; q_x$, corresponding to a decreasing
wavelength, which the shear--advected fluctuation exhibits along the $y$--direction.
Taking into account this time--dependence of the wavelength of fluctuations, a
correlation function characterized by a single wavevector can be defined,
which resembles the equilibrium quantity:
$C_{f_{\bf k} g_{\bf q}}(t) = N C_{fg; {\bf q}}(t) 
\delta_{\vek{q}(t) , \vek{k}}$ 
with:
\beq{b15}
C_{fg ; {\bf q}}(t) = \frac 1N \, \langle
\delta f^*_{\vek{q}(t)}
 \;  e^{\smopb\, t }\,  \delta g_\vek{q} \rangle^{(\gd)}\; , \quad\mbox{with }\; 
{\bf q}(t) = {\bf q} + \vek{q} \kapl\, t\; .
\eeq
Picking out density fluctuations $\rho_{\bf q}(t)$ again, the abbreviation 
$C_{\bf q}(t) = \frac 1N \, \langle \rho^*_{\vek{q}(t)}
 \;  e^{\smopb\, t }\,  \rho_\vek{q} \rangle^{(\gd)}$ for the  intermediate scattering function
under shear will be used.   
Similarly for the susceptibilities from \gl{a4} one finds $\chi_{f_{\bf k} g_{\bf q}}(t) = N \chi_{fg; {\bf q}}(t) 
\delta_{\vek{q}(t) , \vek{k}}$ with the result:
\beq{b16}
\chi_{fg ; {\bf q}}(t) = \frac 1N \, \langle
 \sum_i \; \frac{\partial f_{{\bf q}(t)}^*}{\partial {\bf r}_i} \cdot \pa_i \;
e^{\smopb t}\; g_{\bf q} \rangle^{(\gd)}  \; ,
\eeq
where the specialisation to density variables  shall be denoted by $\chi_{\bf q}(t)= \chi_{\rho\rho ; {\bf q}}(t)$.

\dred{While these expressions are easily formulated, they suffer from a lack of knowledge about
 $\Psi_s$. Thus in the following a formal framework is developed within which to approximate $\Psi_s$.}

\section{Transient dynamics approach}
\label{green}

The following situation shall be studied: The system is in equilibrium at 
times $t\le0$, when instantaneously a constant shear rate $\gd$ is turned on:
\beq{c1}
\smop(\Gamma,t) 
= \left\{ \begin{array}{ll}
\Omega_e(\Gamma) & t\le 0\; ,\\
\smop(\Gamma) & t> 0\; ,
\end{array}\right.
\eeq
so that the distribution function at $t=0$ coincides with the equilibrium one,
$\Psi(\Gamma,t=0)=\Psi_e(\Gamma)$.
The solution of \gl{c1} is easily found for $t\ge0$:
\beq{c2}
\Psi(\Gamma,t) =  e^{\smop(\Gamma)\, t}\, \Psi_e(\Gamma)
\; . 
\eeq
The switching--on of a real rheometer is supposed to influence the initial variation of $\Psi(t)$ only,
 which will be neglected in the following as the stationary state, presumably 
reached for $t\to\infty$, will be  considered. 
Rewriting the exponential function,
\[
e^{\Omega t} = 1 + \int_0^t\!\!\! dt' \; e^{\Omega t'}\; \Omega \; ,
\]
leads together with \gls{a5}{a6} to the formal result for the
steady state distribution function (where physical units are restored, and the adjoint
SO is introduced acting on the variables to be averaged with $\Psi_s$):
\beq{c3}
\Psi_s(\Gamma) = \Psi_e(\Gamma) + \frac{\gd}{k_BT} \; \int_0^\infty\!\!\!\! dt' \; 
\Psi_e(\Gamma) \; \sigma_{xy} \; e^{\smopb(\Gamma) t' } \; .
\eeq
This simple result is central to our approach as it connects steady state properties to time
integrals formed with the shear--dependent dynamics. Knowledge about slow relaxation processes in 
the system can enter.
Consequently, the steady--state averages from \gl{b13} are given by
\beqa{c4}
f(\gd) & = & \langle f_{\bf q=0} \rangle / V + 
\frac{\gd}{V}  \; \int_0^{\infty}\!\!\!\!\! dt' \; 
\langle \sigma_{xy}  \;
e^{ \smopb  t'  }\,  f_{\bf q=0} \rangle
\; ,
\eeqa
while corresponding expressions hold for the structure functions from \gl{b14},
\beqa{c5}
S_{fg; {\bf q}}(\gd) & = &  \langle \delta f^*_{\bf q}\; \delta g_{\bf q}  \rangle / N +
\frac{\gd}{N}  \; \int_0^{\infty} \!\!\!\!\! dt' \; 
\langle \sigma_{xy}  \;
e^{ \smopb  t'  }\,  \delta f^*_{\bf q}\; \delta g_{\bf q} \rangle
\; ,
\eeqa
the fluctuation functions from \gl{b15},
\beqa{c6}
\hspace*{-0.6cm}C_{fg; {\bf q}}(t) &  = & \langle \delta f^*_{{\bf q}(t)} \; 
 e^{\smopb\, t }\, \delta g_{\bf q} \rangle / N  + \frac{\gd}{N}   \int_0^\infty\!\!\!\!\! dt' \;
\langle  \sigma_{xy} \; e^{\smopb\, t' }\,  \delta f^*_{{\bf q}(t)}\; 
 e^{\smopb\, t }\, \delta g_{\bf q}  \rangle \, ,
\eeqa
and the susceptibilities from \gl{b16} 
\beqa{c7}
\hspace*{-0.3cm}\chi_{fg; {\bf q}}(t) &  = & - \langle \delta f^*_{{\bf q}(t)} \; \smopb_e\; 
 e^{\smopb\, t }\, \delta g_{\bf q} \rangle / N  \nonumber \\ & & - \frac{\gd}{N}   \int_0^\infty\!\!\!\!\! dt' \;
\langle  \sum_i \left( {\bf F}_i + \pa_i \right) \sigma_{xy} \; e^{\smopb\, t' }\; \delta f_{{\bf q}(t)}^*  \; \pa_i \;
 e^{\smopb\, t }\, \delta g_{\bf q}  \rangle \; .
\eeqa
Note that the averages in \glto{c4}{c7} can be performed with the known equilibrium distribution function. When studying the nonlinear rheology of simple fluids, 
transient correlation functions related to \gl{c4} were found useful in thermostatted simulations \cite{Morriss87} and in mode coupling approaches \cite{Kawasaki73b}.

\subsection{Translational invariance reconsidered}
\label{trans}

The time--dependent distribution function $\Psi(\Gamma,t)$ from \gl{c2} can be used to show that a translationally invariant equilibrium distribution function $\Psi_e(\Gamma)$ leads to a translationally invariant steady state distribution $\Psi_s(\Gamma)$.
To that end, as in Sect. {\it \ref{tra}}, $\Psi(\Gamma,t)$ is considered at the shifted positions, $\Gamma \to \Gamma'$ with ${\bf r'}_i={\bf r}_i + {\bf a}$ for all $i$:
\beq{y1}
\Psi(\Gamma',t) =  e^{\smop(\Gamma) \, t - {\bf P} \, \kap \, {\bf a} \, t} \; \Psi_e(\Gamma) \; ,
\eeq
where $\Psi_e(\Gamma')=\Psi_e(\Gamma)$ was used. The SO $\smop$ and the operator ${\bf P}\,\kap\,{\bf a}$ with ${\bf P}$ from \gl{b115}
commute, because the shear rate tensor satisfies $\kapa\cdot \kapa = 0$, and because the sum of all internal forces vanishes due to Newton's third law:
\beqa{y2}&  
\left( {\bf P} \kap {\bf a} \right) \; \smop - \smop \; \left( {\bf P} \kap {\bf a} \right)  = & \nonumber \\ 
&  \sum_{j} \left\{ \left[ \pa_j ( \cdot \frac{\partial}{\partial {\bf r}_j} 
\left( \sum_i \frac{\partial U}{\partial {\bf r}_i}\kap{\bf a} \right) ) \right] -  
\left[ ({\bf a}\kapl^T  \cdot \kaprt\pa_j) \right] \right\} = 0
\; . & 
\eeqa
Therefore, the Baker--Hausdorff theorem simplifies \gl{y1} to
\beqa{y3}
\Psi(\Gamma',t) &=&  e^{\smop(\Gamma) \, t }\; e^{- {\bf P} \, \kap \, {\bf a} \, t} \; \Psi_e(\Gamma)\nonumber \\
 &=& e^{\smop(\Gamma) \, t }\; e^{- \left( \sum_i {\bf F}_i \right) \, \kap \, {\bf a} \, t} \; \Psi_e(\Gamma) \nonumber\\
 &=& e^{\smop(\Gamma) \, t } \; \Psi_e(\Gamma) \; ,
\eeqa
where the last equality again holds because the sum of all internal forces vanishes. Therefore, 
\beq{y4} 
\Psi(\Gamma',t) = \Psi(\Gamma,t) 
\eeq
holds, proving that the time--dependent and consequently the stationary distribution function $\Psi_s(\Gamma)=\lim_{t\to\infty} \Psi(\Gamma,t)$
\dred{are translationally invariant even though the SO from \gl{a1} itself is not. 
This applies, at least, in cases without spontaneous symmetry breaking. Formally, the role of such symmetry breaking is to discard some parts of the steady state distribution function and keep others (with the choice dependent on initial conditions). The distributions developed here discard nothing, and would therefore average over the disjoint symmetry-related states of a symmetry-broken system.}

\subsection{Particle conservation}

For the present approach to be tenable, the time--dependent distribution
function in \gl{c2} needs to approach $\Psi_s$ at long times. Putting  aside at first questions of non--ergodic glassy dynamics, the consequences of the conserved variables in the system need to be addressed. The particle number is the only conserved quantity, and its microscopic conservation law reads from \gl{b11}:
\beq{c8}
\partial_t \rho_{\bf q}(t) =  \smopb \; \rho_{\bf q}(t) = i {q}  \; {j}^{||}_{\bf q}(t) \; ,
\eeq
with the particle flux given by the longitudinal element of the stress tensor (this holds generally for overdamped motion as the velocity is proportional to the force) and
the drift flux, ${j}^{||}_{\bf q}(t) = - i q\; \sigma^{||}_{\bf q}(t) - i\, v^{\kapa}_{\bf q}(t)$, with:
\beqa{c9}
\sigma^{||}_{\bf q}(t) &=& - e^{\smopb t}\; \sum_i \; \left( 1 - \frac{i}{q^2}\; {\bf q \cdot F}_i
\right) \; e^{i {\bf q}\cdot{\bf r}_i} \nonumber\\
 v^{\kapa}_{\bf q}(t) &=&  e^{\smopb t}\; \sum_i 
\; \frac{i{\bf q}\kap{\bf r}_i}{q}   
\; e^{i {\bf q}\cdot{\bf r}_i} \; .
\eeqa

To verify that particle number conservation does not prevent decay of the
dynamics in \glto{c3}{c7}, the slow density fluctuations are eliminated 
using the equilibrium projection operator \cite{dhont}: 
\beq{c10}
P = \rho_\vek{q} \rangle (NS_q)^{-1} \langle \rho^*_\vek{q} \wer
P A = \rho_\vek{q} \; \frac{\langle \rho^*_\vek{q} A \rangle}{(N S_q)}  \; ,
\eeq
with complement $Q=1-P$, where sums over wavevectors $q$ are suppressed 
because of orthogonality. Here, the equilibrium static structure factor is abbreviated as
$S_q=S_{\bf q}(\gd=0)=\langle \rho_{\bf q}^*\,\rho_{\bf q}\rangle/N$ (it will appear repeatedly in the following), and idempotency  $P^2 = P$ is clear.

The correlation functions in \glto{c4}{c7} can be abbreviated by
$\langle \sigma_{xy}\; e^{\smopb t}\; X \rangle$,  with  $X=f_{\bf 0} $ in \gl{c4}, 
$X=\delta f^*_{\bf q}\, \delta g_{\bf q}$ in \gl{c5}, and $X=\delta f^*_{{\bf q}(t')}\,e^{\smopb t'}\, \delta g_{\bf q}$ in \gl{c6};  employing the projector $P$ and recalling $\langle
\sigma_{xy} \rho_{\bf q} \rangle = 0$, these become:
\beq{c11}
\langle \sigma_{xy} e^{\smopb t} X \rangle = 
\langle \sigma_{xy} \, Q\; e^{\smopb t} \; Q \: X \rangle
+ \langle \sigma_{xy} \; e^{\smopb t} \; \rho_{\bf q} \rangle \; \frac{\langle \rho_{\bf q}^*\;
X \rangle}{N S_q}\; .
\eeq
In the first term on the right hand side, already only fluctuations which are perpendicular to the hydrodynamic densities appear. The projector \gl{c10} can be used to show the vanishing of the second term, where the slow hydrodynamic  modes could enter. From \gl{c12} in the Appendix $\langle \sigma_{xy} \; e^{\smopb t} \; \rho_{\bf q} \rangle =0$ follows, as does the fact that reduced dynamics and full dynamics agree for the correlation functions needed in \glto{c4}{c7}, viz.:
\beq{c14}
\langle \sigma_{xy} \; e^{\smopb t} \; X \rangle  =\langle \sigma_{xy} \; Q \; e^{\smopb t} \;Q\; X \rangle  =
\langle \sigma_{xy} \; Q \; e^{Q \, \smopb \, Q \, t} \; Q \; X \rangle  \; .
\eeq
The result is perhaps not surprising. The fact that
density fluctuations are independent of the applied velocity field causes the dynamics 
leading to the  changes in the stationary expectation values to be orthogonal
to linear density fluctuations. 

\subsection{Generalized Green--Kubo relations}

The result \gl{c14}  obtained from considering the conserved density fluctuations completes our derivation of
{\em generalized Green--Kubo relations}. While the familiar Green--Kubo relations of linear response theory connect transport coefficients to time--integrals over projected fluxes \cite{Forster75}, \glto{c4}{c7} enable one to calculate the steady state properties of strongly sheared states far from  equilibrium. Because of \gl{c14}, the conserved density fluctuations do not contribute and the dynamics contains no hydrodynamically slow mode.

\section{Transient density fluctuations}
\label{tradens}

\dred{The problem of calculating  steady state  averages is thus converted into one of first finding the transient dynamics after switching on the rheometer,
and then integrating this in order  to use \glto{c4}{c7}.} 
The transient density fluctuations will be important in this \dred{process} (see Ref. \cite{Fuchs02}) and thus shall be simplified first. 
Because of the equivalence of the particles, the normalized transient collective intermediate scattering function  can be written as:
\beq{e1}
\Phi_\vek{q}(t) = \frac{1}{N S_q}\; \langle \rho_{{\bf q}(t)}^* \;
 e^{\smopb t} \rho_\vek{q} \rangle = \frac{1}{S_q}\; 
\langle \rho_\vek{q}^{s*} \, e^{- i \vek{q} \kap \vek{r}_s
t } \; e^{\smopb t} \rho_\vek{q} \rangle \; ,
\eeq
where $\rho_\vek{q}^s=e^{i \vek{q}\cdot{\bf r}_s}$ is the density of a single tagged particle, which is identical to the others.  
By this trick of singling out a particle, the motion of the surrounding particles due to the imprinted flow profile
can be specified exactly in the time evolution described with one SO. By differentiating one finds:
\[ \hspace*{-1cm}
\partial_t \;
e^{ - i \vek{q} \kap \vek{r}_s t } 
\; e^{\smopb t} = 
( - i \vek{q} \kap \vek{r}_s 
+  e^{ - i \vek{q} \kap \vek{r}_s t } \; \smopb \;
 e^{ i \vek{q} \kap \vek{r}_s t } 
) \; e^{ - i \vek{q} \kap \vek{r}_s t } \;  e^{\smopb t} \; . \]
Integrating in time, the time evolution operator incorporating advection, in the
case of density fluctuations, is found as:
\beqa{e2}
\Phi_\vek{q}(t) &=& \frac{1}{S_q}\; \langle \rho_\vek{q}^{s*} \; e_+^{\int_0^t d\tau\, 
\Omega_s(\tau)}  \;
\rho_\vek{q} \rangle \; , \wer \nonumber\\  \Omega_s(t) &=& 
 - i \vek{q} \kap \vek{r}_s    +  e^{ - i \vek{q} \kap \vek{r}_s t } 
\, \smopb \,  e^{ i \vek{q} \kap \vek{r}_s t }   \; ,
\eeqa
and $e_+$ is the time ordered exponential, where earlier times appear on the 
right. The time evolution operator  can be worked out explicitly (but to little avail):
\beq{e3}
\Omega_s(t) = 
\smopb - i \vek{q} \kap \vek{r}_s + i  \vek{q} \kap \left( 2 \pa_s  + 
 \vek{F}_s \right) \, t - \vek{q} \kap \kaprt \vek{q} \, t^2 \; ,
\eeq
and because of shear--advection is explicitly time--dependent.

Projection operator manipulations \cite{Kawasaki73b} simplify the time ordered exponentials, see the Appendix, and from \gl{z1} follows \dred{an exact} Zwanzig--Mori type equation of motion for the transient density correlators: 
\beq{e4}
 \partial_t \Phi_\vek{q}(t) + \Gamma_\vek{q}(t)\, 
\Phi_\vek{q}(t) +  \int_0^t dt'\; M_\vek{q}(t,t') \;  \Phi_\vek{q}(t') = 0 \; .
\eeq
Here the static  projector $P_s = \rho_\vek{q} \rangle (1/S_q) \langle \rho_\vek{q}^{s*}$
was employed; because of the equivalence of the particles it again satisfies $P_s^2=P_s$. The time--dependence of $\Omega_s(t)$ leads to a slightly more general time--dependence in \gl{e4} than familiar in equilibrium.  The `initial decay rate' from \gl{z2} is
\beqa{e5}
 \Gamma_\vek{q}(t) &=&  - \frac{\langle \rho^{s*}_\vek{q}  \; \Omega_s(t) \;
\rho_\vek{q} \rangle}{S_q} \nonumber \\  &=&
 \frac{ q^2 + q_x q_y \; \gd t }{S_q} + 
( q_x q_y \;\gd t + q_x^2 \; \gd^2 t^2 ) - \frac{q_x q_y }{q S_q}\; \gd\;  \frac{\partial S_q}{\partial q} \; ,
\eeqa
which recovers the `Taylor dispersion' familiar for non--interacting particles 
\cite{dhont};  for non--interacting particles $M_\vek{q}(t,t')=0$ holds.

The memory function in \gl{e4} is given by
\beqa{e6}
M_\vek{q}(t,t') &=& -
\langle A_\vek{q}^{s*}(t) \; U_s(t,t') \; B^s_\vek{q}(t') \rangle/ S_q 
\; , \wer \nonumber\\
U_s(t,t') &=& 
e_+^{\int_{t'}^{t} d\tau\; \Omega_s(\tau) \; Q_s } \; ,
\eeqa
and where the generalized longitudinal stress tensor elements are 
\beqa{e7}
 \langle A_\vek{q}^{s*}(t) &=& \langle \rho^{s*}_\vek{q} \; \Omega_s(t) Q_s 
 \nonumber\\ 
B^s_\vek{q}(t)\rangle  &=& \Omega_s(t) \; \rho_\vek{q} \rangle \; ,
\eeqa
see \gl{z1}.
The memory function $M_\vek{q}(t,t')$ encodes the after--effects of the variables not treated explicitly in $\Phi_{\bf q}(t)$ that provide a bath for the density fluctuations. 

\dred{In the context of mode-coupling theory, much depends on how this intractable object is approximated.} There is ample evidence, for dense colloidal dispersions close to equilibrium, that a Markovian approximation for $M$ is quite insufficient \cite{Pusey91}. Also evidence from careful dynamic light
scattering tests of mode coupling theory close to equilibrium \cite{Megen93,Megen94b} suggest that following Cichocki and Hess  \cite{Cichocki87} a second projection step is required. \dred{We perform this step now; further discussion is deferred to} Sect. {\it \ref{discmM}} below. 

In the second projection operator step, the time evolution operator, \gl{e3}, is formally \cite{Cichocki87,Kawasaki95} 
decomposed as
\beqa{e8}
\Omega_s(t) &=& \Omega_s^i(t) + \Omega_s(t) \; \rho_\vek{q} \rangle \; 
\langle \rho_\vek{q}^{s*}\; \Omega_s(t)\; \rho_\vek{q} \rangle^{-1} \; \langle
\rho_\vek{q}^{s*}\; \Omega_s(t) \nonumber \\ &=& \Omega_s^i(t) + \Omega_s^{\rm red.}(t)  \; , 
\eeqa
where the `reducible'  part of the SO couples the dynamics back to the generalized stress elements:
\beq{e85}
 \Omega_s^{\rm red.}(t)\;  Q_s = - B^s_\vek{q}(t) \rangle \; \frac{1}{S_q \;
 \Gamma_\vek{q}(t) } \;  \langle A^{s*}_\vek{q}(t) \; .
\eeq
The importance of this separation of $\Omega_s(t)$ lies in the possibility to introduce another memory function.
While $M_\vek{q}(t,t')$ plays the role of a generalized diffusion kernel, the new memory function $m_\vek{q}(t,t')$ plays the role of a generalized friction kernel. 
As shown in the Appendix, see \gl{z3}, the original memory function can  be rewritten using \gl{e8} as
\beqa{e9} 
M_\vek{q}(t,t') &+& \Gamma_\vek{q}(t) \; m_\vek{q}(t,t') \; \Gamma_\vek{q}(t')  \nonumber \\  &+& 
\Gamma_\vek{q}(t) \int_{t'}^t\!\!\! dt''\; 
 m_\vek{q}(t,t'') \; M_\vek{q}(t'',t')  = 0  \; ,
\eeqa
where the new memory function is defined as:
\beq{e10}
m_\vek{q}(t,t') =  \Gamma_\vek{q}^{-1}(t) \;
\langle A^{s*}_\vek{q}(t) \; U_s^i(t,t') \; B^s_\vek{q}(t') \rangle 
\; \Gamma_\vek{q}^{-1}(t')\; /S_q \; .
\eeq
Its time--dependence is given by the `irreducible' \cite{Cichocki87,Kawasaki95} dynamics introduced in \gl{e8}:
 \beq{e105}
U_s^i(t,t') = e_+^{\int_{t'}^{t} d\tau\; \Omega_s^i(\tau) Q_s } 
\; .
\eeq
From the theory of Volterra integral equations \cite{Tricomi57}, see \gl{z4} in the Appendix,
it \dred{follows} that the equation  of motion, \gl{e4}, thus can be rewritten as:
\beq{e11}
 \partial_t \Phi_\vek{q}(t) + \Gamma_\vek{q}(t) \; \left\{
\Phi_\vek{q}(t) + 
 \int_0^t dt'\; m_\vek{q}(t,t') \; \partial_{t'}\,  \Phi_\vek{q}(t')
\right\} = 0 \; .
\eeq
Equation (\ref{e11}), together with the definition of the memory function $m$ in \gl{e10},
is the central new result of the approach to shear thinning introduced in Ref. 
\cite{Fuchs02}, \dred{and is derived explicitly here for the first time}. Together with the generalized Green--Kubo relations of \glto{c4}{c7}, it will be the starting point for factorizations building
on the insights of mode coupling theory \cite{Goetze91b} into the dynamics of quiescent colloidal dispersions.

\subsection{Discussion of the memory functions $M$ and $m$}
\label{discmM}

The equations of motion containing the two memory functions differ because of the
Cichocki--Hess projection step in \gl{e8}.  
The different contents of \gls{e4}{e11} can be seen from performing
a Markovian approximation in the two memory functions. Then, \gl{e4} becomes
\beq{e12}
 \partial_t \Phi_\vek{q}(t) + \left[ \Gamma_\vek{q}(t) +
 \int_0^t dt'\; M_\vek{q}(t,t')  \right]\;  \Phi_\vek{q}(t) 
= 0 \;, 
\eeq
where the renormalization of the decay rate $\Gamma$ 
can be expected to be negative 
(without shear this can be shown rigorously, and is connected to \gl{a55}).
 In order to describe a
 slowing down of the dynamics (viz. a small effective decay rate) thus a near cancellation of the two terms in the 
 square bracket in \gl{e12} is required. Approximations will need to be subtle to recover this near cancellation.

On the other hand, \gl{e11} becomes upon performing a Markovian approximation:
\beq{e13}
 \partial_t \Phi_\vek{q}(t) + \left[ \Gamma_\vek{q}(t) \left/ 
\left( 1 + \Gamma_\vek{q}(t)  \int_0^t dt'\; m_\vek{q}(t,t')  \right) \right. \right]\;  
\Phi_\vek{q}(t) = 0 \;, 
\eeq
where the renormalization of the decay rate now describes a suppression  
of the dynamics (slowing down) as the memory contribution can be expected to
be positive (again without shear rate this can be shown rigorously), and large. 
Any approximation yielding a large memory--integral
 (as expected close to equilibrium in dense dispersions), 
thus can reasonably describe slowing down using \gl{e11} without running the risk to predict
an unstable system, viz.  negative decay rates. 

\subsection{Neutral or vorticity direction}

In the vorticity direction, $\vek{q}=q\, \hat{\bf z} $, perpendicular to the impressed solvent flow and its gradient, 
the \gls{e10}{e11} simplify to almost the known ones from the standard
Zwanzig--Mori approach. The equation of motion becomes
\beq{e14} 
 \partial_t \Phi_{q\hat{\bf z} }(t) + \frac{q^2}{S_q} \left\{ \,
\Phi_{q\hat{\bf z} }(t) + 
 \int_0^t dt'\; m_{q\hat{\bf z} }(t,t') \; \partial_{t'}\,  
\Phi_{q\hat{\bf z} }(t') \right\} = 0 \; ,
\eeq
with the simpler expression of the memory function:
\beqa{e15}
m_{q\hat{\bf z}}(t,t') &=& \langle \left( \sigma^{||*}_{q\hat{\bf z}} +
V^{\kapa *}_{q\hat{\bf z}} \right)\, Q \; e^{\Omega^i_z 
(t-t')}\;  \sigma^{||}_{q\hat{\bf z}}  
\rangle \; S_q/N \; , \wer \nonumber \\ V^{\kapa}_{q\hat{\bf z}} &=&  
 \sum_{jl} e^{ i q z_l}\; \frac{\vek{r}_j
\kapt \vek{F}_j}{q^2} \; Q \; .
\eeqa
The stress tensor $\sigma^{||}_{q}$  was defined in \gl{c9}.
To simplify $m$ the equivalence of the particles was used, replacing the single particle fluctuation
$\rho_{\bf q }^s$ by the collective one $\rho_{\bf q }/N$ in all averages where the index $s$ appears only once. 
Importantly, shearing  affects the vorticity direction not only via $V^{\kapa}$, but also via the reduced dynamics which 
couples all spatial directions and contains the shear rate $\gd$ in any order:
\[\Omega^i_z = \smopb \; Q +   \sigma^{||}_{q\hat{\bf z}}  
\rangle \; \frac{1}{q^2} \; \langle  \left( \sigma^{||*}_{q\hat{\bf z}} +
V^{\kapa *}_{q\hat{\bf z}} \right) \, Q \; .
\]
Besides the reassurance that the formal manipulations recover almost standard results in the case where shearing
affects the particle motion least, the result \gl{e15} for the memory function in the vorticity direction is noteworthy for two reasons: First, the stress--stress autocorrelation function $m$ calculated from \gl{a1} without shear arises from potential interactions and thus approaches a constant for vanishing wavevector, $m_q(t) \to m_0(t)<\infty$ for $q\to0$ \cite{Pusey91}. With shear, however, the result that $q^2\, m_{q\hat{\bf z}}(t,t') \to $ const. for $q\to0$ can be expected from \gl{e15}, because the particles are forced by the flow field. Thus the hydrodynamic collective diffusion process will be affected. Second, the complicated time dependence which arises in the memory function $m_{\bf q	}(t,t')$ in $x$-- and $y$--direction because of the advection of stress fluctuations with the imposed flow, simplifies to a dependence solely on the {\em time--difference} between stress fluctuations along the vorticity direction, $m_{q\hat{\bf z}}(t,t')=  m_{q\hat{\bf z}}(t-t')$.

\section{Discussion and outlook}
\label{disc}

The derived generalized Green--Kubo relations of \glto{c4}{c7}, and the equations of motion for the transient density correlators \gl{e11} follow from the Smoluchowski equation for Brownian particles under \dred{uniform imposed shear as given} in \gl{a1}. From the Green--Kubo relations, general conclusions about linear response around the equilibrium state can be made by setting $\gd=0$ in the dynamics. The reduced dynamics in \gl{c14}  contains no hydrodynamic components, because the density is the only conserved
variable in Brownian systems. As long as the equilibrium fluctuations are 
ergodic and decay faster than $1/t$ for long times, the leading change in any stationary variable
is linear in shear rate $\gd$. Non--linearities in $\gd$ in steady state quantities will be largest for variables where the transient dynamics exhibits the slowest algebraic decay. 

A central approximation of the approach is hidden in our postulate that the
time--dependent solutions to the Smoluchowski equation approach the stationary solution at long times. Aging effects \cite{Latz02} could prevent glassy quiescent states to follow the transient dynamics calculated above. Spatial symmetry breaking could lead to inhomogeneous states, such as 'shear--banded' ones. Only comparison with simulations and  experiments can determine whether systems exist exhibiting the postulated properties.

A central difference to standard equilibrium Zwanzig--Mori equations for density fluctuations is the appearance of the
time--dependent {\em wavevector advection} in the time--evolution operator $\Omega_s(t)$ of \gl{e2}. It arises because of the affine deformation of fluctuations and is an exact consequence of shear in the Smoluchowski equation of \gl{a1}. As found for simple liquids \cite{Miyazaki02,Miyazaki04}, the mode coupling approximations described in Refs. \cite{Fuchs02,Fuchs03,Fuchs03b} deduce from it that shear speeds up the structural relaxation and thus causes shear--thinning.  
The aspect that the stationary nonequilibrium state is characterized by a non--vanishing probability current, which is connected to the non--Hermitian nature of the Smoluchowski operator, enters our approach in the strategy to calculate the steady state distribution function via integrating through the transient.

As discussed in Sect. {\it \ref{discmM}}, the memory function $m_{\bf q}(t,t')$ from \gl{e11} and the equation of motion 
\gl{e10} appear reasonable starting points for approximations capturing the slow dynamics in driven (sheared) dense colloidal dispersions. Mode coupling approximations had been suggested in \cite{Fuchs02}, their universal contents had been discussed in \cite{Fuchs03}, and their detailed presentation will be given in a future companion publication.

 \ack

We thank M. Ballauff,  J. Bergenholtz, Th. Franosch, 
A. Latz, K. Kroy, and G. Petekidis for discussions, and Th. Franosch for a critical reading of the manuscript. 

\appendix

\section{}
The appendix contains various more technical manipulations, which are used in the main text.

\subsection{}

The calculation of the linear response susceptibility in the stationary state starts from the change in the energy given in \gl{aa6}. The SO changes to $\smop-\Delta \Omega(\Gamma,t)$, where
\beq{ap1}
\Delta \Omega(\Gamma,t) = \sum_i\; \pa_i \cdot
\left( \frac{\partial}{\partial {\bf r}_i} \, f^*(\Gamma) \right) \; h_e(t) \; .
\eeq
To linear order in the external field $h_e$ the stationary distribution function changes to:
\beq{ap2}
\Psi(\Gamma,t) = \Psi_s(\Gamma) - \int_{-\infty}^t\!\!\!\!dt'\; e^{\smop\, (t-t')}\; \Delta \Omega(\Gamma,t')\; \Psi_s(\Gamma) + {\cal O}(h_e^2) \; ,
\eeq
which leads to the shift of an arbitrary expectation value linear in the external field $h_e$ given by:
\beq{ap3}\hspace*{-0.5cm}
\hspace*{-0.3cm}\langle g \rangle^{(\gd , h_e)} - \langle g \rangle^{(\gd)} = -  \int_{-\infty}^t\!\!\!\!dt'\; g(\Gamma)\; e^{\smop\, (t-t')}\; \Delta \Omega(\Gamma\,t')\; \Psi_s(\Gamma) + {\cal O}(h_e^2) \; .
\eeq
A partial integration leads to \gl{aa5} with the definition of the susceptibility in \gl{a4}.

\subsection{}

In order to show that conserved density fluctuations do not prevent the dynamics in \glto{c4}{c7} from relaxing,
 the following operator equality is useful, where the first line can be shown straightforwardly by
 differentiation: 
\beqa{c12}
e^{\smopb t} &=&  e^{\smopb\, Q t} + \int_0^t \!\!\! dt'\;  
e^{\smopb t'} \; \smopb\, P \; e^{\smopb\, Q\, (t-t')} \nonumber \\
&=& e^{\smopb\, Q t} + \frac{i q}{N S_q} \; \int_0^t \!\!\! dt'\;  
e^{\smopb t'} \; j^{||}_{\bf q} \rangle\; \langle \rho_{\bf q}^* \; e^{\smopb\, Q\, (t-t')}  \; ,
\eeqa
where $P$ is the projection operator from \gl{c10}.
Two conclusions can be drawn. First, because $\langle \sigma_{xy} \; e^{\smopb t} \; \rho_{\bf q} \rangle$ in \gl{c11} by translational symmetry (see \gl{b12}) can be non--vanishing for ${\bf q}=0$ only, and there only the first term in \gl{c12} survives, it follows that:
\beqa{c13} 
\langle \sigma_{xy} \; e^{\smopb t} \; \rho_{\bf q} \rangle &=& 
\langle \sigma_{xy} \; e^{\smopb\, Q\, t} \; \rho_{\bf q} \rangle \; \delta_{\bf q , 0}
\nonumber\\
&=& \langle \sigma_{xy} \; \rho_{\bf q} \rangle \; \delta_{\bf q , 0} = 0 \; ,\nonumber
\eeqa
as can be seen by expanding the exponential and using $Q \, \rho_{\bf q} =0$. Consequently,  
\gl{c11} simplifies to 
\[
\langle \sigma_{xy} \; e^{\smopb t} \; X \rangle  =\langle \sigma_{xy} \; Q \; e^{\smopb t} \;Q\; X \rangle \; .\]
 Second,  reduced dynamics and full dynamics agree for the correlation functions needed in \glto{c4}{c7}, viz.:
\beq{c145}
\langle \sigma_{xy} \; Q \; e^{\smopb t} \;Q\; X \rangle  =
\langle \sigma_{xy} \; Q \; e^{\smopb \, Q \, t} \; Q \; X \rangle  \; ,
\eeq
again, because of the vanishing of the difference (arising from the second term on the last line in \gl{c12} )
 at $q=0$. This result leads to \gl{c14}.

 \subsection{}
 
In order to derive the equation of motion for transient density fluctuations, the time ordered product in \gl{e2} is
rewritten using the projection operator $Q_s$, which is the complement (viz. $1=P_s+Q_s$) to $P_s$ defined below
  \gl{e4}:
 \beq{zz1}
  e_+^{\int_0^t d\tau\,  \Omega_s(\tau)}  =  U_s(t,0) +
\int_0^t \!\!\! ds\;  U_s(t,s) \; \Omega_s(s) \, P_s \: 
e_+^{\int_0^s d\tau\,  \Omega_s(\tau)} \; ,
\eeq
with the abbreviation $U_s(t,t')$ from \gl{e6}. Equality can be shown by differentiation:
 \beqa{z0}
\partial_t\,  e_+^{\int_0^t d\tau\,  \Omega_s(\tau)}  &=&
\Omega_s(t) \, Q_s \;  U_s(t,0)  +
\Omega_s(t) \, P_s \; e_+^{\int_0^t d\tau\,  \Omega_s(\tau)}  \nonumber \\ & & +
\int_0^t \!\!\! ds\; \Omega_s(t) \, Q_s \; U_s(t,s) \; \Omega_s(s) \, P_s \: 
e_+^{\int_0^s d\tau\,  \Omega_s(\tau)} \; , 
\eeqa
where $\partial_t\, U_s(t,t') = \Omega_s(t) \, Q_s U_s(t,t')$ was used.
Regrouping on the right hand side, shows that $(\partial_t - \Omega_s(t) Q_s )\, \eta(t)= 
\Omega_s(t) \, P_s
\,  \exp_+{\int_0^t d\tau\,  \Omega_s(\tau)}$, where $\eta(t)$ abbreviates either the left or right hand side of
\gl{zz1}. Yet, \gl{z0} turns out more useful when sandwiched between density fluctuations:
\beqa{z1} &\hspace*{-1cm}
\partial_t \, \Phi_\vek{q}(t)\, S_q = 
\langle \rho_\vek{q}^{s*} \; \partial_t\,  e_+^{\int_0^t d\tau\,  \Omega_s(\tau)} \; \rho_\vek{q} \rangle  =
  \frac{\langle \rho_\vek{q}^{s*} \Omega_s(t) \rho_\vek{q} \rangle}{S_q} \;  \langle \rho_\vek{q}^{s*} \;  e_+^{\int_0^t d\tau\,  \Omega_s(\tau)} \; \rho_\vek{q} \rangle & \nonumber \\ &  +
\int_0^t \!\!\! ds\; \langle \rho_\vek{q}^{s*} \Omega_s(t) \, Q_s \; U_s(t,s) \; \Omega_s(s) \,\rho_\vek{q} \rangle \frac{1}{S_q}  \langle \rho_\vek{q}^{s*}  \: 
e_+^{\int_0^s d\tau\,  \Omega_s(\tau)} \rho_\vek{q} \rangle \; .&
\eeqa 
Here, $Q_s \, U_s(t,0)\, \rho_{\bf q} = Q_s \, \rho_{\bf q} = 0$ was used.
Equation (\ref{e4}) follows and the definitions of the rate $\Gamma_{\bf q}(t)$ and the  memory function $M_{\bf q}(t,t')$. 
The rate $\Gamma_{\bf q}(t)=-S_q \langle \rho^{s*}_\vek{q}  \; \Omega_s(t) \; \rho_\vek{q} \rangle$
 can easily be evaluated, using \gl{e3}:
\beqa{z2}
& \langle \rho^{s*}_\vek{q}  \; \Omega_s(t) \;
\rho_\vek{q} \rangle = \langle \rho^{s*}_\vek{q}  \; \left( \Omega_e - \vek{q} \kapl \kapt \vek{q} \, t^2 \right) 
\; \rho_\vek{q} \rangle & \nonumber \\
&   + i \vek{q} \kap \, \langle \rho^{s*}_\vek{q}  \;  \left( 2 \pa_s + {\bf F}_s \right) \; \rho_\vek{q} \rangle 
+ \sum_i \langle \rho^{s*}_\vek{q}  \;  \left( \vek{r}_i \kapt \pa_i - \vek{q} \kap {\bf r}_s 
\delta_{i , s} \right) \; \rho_\vek{q} \rangle &
\nonumber\\
& \hspace*{-1cm}
= - q^2\! -\! \vek{q}\kapl\kapt\vek{q} t^2 S_q\! -  \vek{q} \kap \vek{q}  t ( 1 + S_q ) +\!
\sum_i i \vek{q} \kap  \langle \rho^{s*}_\vek{q}\left(\vek{r}_i\!\!\! -\!\!\! {\bf r}_s \right) e^{i \vek{q r}_i} \rangle  \; ,&
\eeqa
which leads to the stated result, because the last term in \gl{z2} becomes  $\vek{q} \kap \frac{\partial}{\partial {\bf q}} S_q$.

\subsection{}

The decomposition of $\Omega_s(t)$ in \gl{e8} leads to the differential equation for the reduced dynamics:
\[\hspace*{-0.5cm}
\partial_t U_s(t,t') = \Omega_s(t) Q_s \, U_s(t,t') = 
\Omega_s^i(t) Q_s\, U_s(t,t') 
+ \Omega_s^{\rm red.}(t) Q_s 
\;  U_s(t,t') \; , \]
which can be viewed as differential equation with $\Omega_s^{\rm red.}(t) Q_s 
\;  U_s(t,t')$ as inhomogeneity. It can be integrated to give
\beqa{z3}
U_s(t,t') &=& U_s^i(t,t') \nonumber \\ & & - \int_{t'}^t dt'' \; U_s^i(t,t'') \;
B^s_\vek{q}(t'') \rangle \; \frac{1}{S_q}\, \Gamma_\vek{q}^{-1}(t'') \; 
\langle A^{s*}_\vek{q}(t'')\;  U_s(t'',t') \; , \nonumber
\eeqa
where the explicit expression for the reducible part from \gl{e85} was used, and
the `irreducible' fast dynamics $U^i_s(t,t')$ 
corresponds to the solution of the homogeneous equation. It is given  in \gl{e105}.
Inserting the expression for $U_s(t,t')$  into the definition of $M_{\bf q}(t,t')$ in \gl{e6} immediately
gives \gl{e9} with the definition of the memory function \gl{e10}.
The equation of motion \gl{e4} can be viewed as a Volterra integral equation of second kind for $\Phi_{\bf q}(t)$,
with kernel proportional to $M_{\bf q}(t,t')$ and $-\partial_t\, \Phi_{\bf q}(t)/\Gamma_\vek{q}(t)$ as inhomogeneity:
\beq{z4}
\Phi_\vek{q}(t) +  \int_0^t dt'\; \frac{1}{\Gamma_\vek{q}(t)}M_\vek{q}(t,t') \;  \Phi_\vek{q}(t') = -
\frac{1}{\Gamma_\vek{q}(t)}\; \partial_t \Phi_\vek{q}(t)   \; .
\eeq
The solution is given by:
\beq{z5}\hspace*{-1cm}
\Phi_\vek{q}(t) = - \frac{1}{\Gamma_\vek{q}(t)}\; \partial_t \Phi_\vek{q}(t) -
  \int_0^t dt'\; \tilde{m}_\vek{q}(t,t') \; \Gamma_\vek{q}(t')\;  \frac{1}{\Gamma_\vek{q}(t')}\;
\partial_{t'}  \Phi_\vek{q}(t') \; ,
\eeq
where the resolvent kernel $\tilde{m}_\vek{q}(t,t')$ satisfies the integral equation \cite{Tricomi57}:
 \beqa{z6} 
& \frac{1}{\Gamma_\vek{q}(t)}\;  M_\vek{q}(t,t') +  \tilde{m}_\vek{q}(t,t') \; \Gamma_\vek{q}(t') & \nonumber \\  & + 
\int_{t'}^t\!\!\! dt''\; 
 \tilde{m}_\vek{q}(t,t'')\, \Gamma_\vek{q}(t'') \, \frac{1}{\Gamma_\vek{q}(t'')}  \; M_\vek{q}(t'',t')  = 0 & \; ,
\eeqa
By comparison of \gls{z6}{e9}, the memory function $m_\vek{q}(t,t')$ of \gl{e10} is identified to agree with the resolvent kernel entering in \gl{z5}, $\tilde{m}_\vek{q}(t,t')=m_\vek{q}(t,t')$. Thus \gl{z5} actually is equivalent to \gl{e11} as was to be shown.

\section*{References}

\begin{thebibliography}{10}

\bibitem{russel}
Russel W~B, Saville D~A  and Schowalter W~R
\newblock  1989 {\em Colloidal Dispersions}
\newblock (Cambridge University Press, New York)

\bibitem{dhont}
Dhont J K~G
\newblock  1996 {\em An introduction to dynamics of colloids}
\newblock (Elsevier Science, Amsterdam)

\bibitem{Risken}
Risken H
\newblock  1989 {\em The Fokker--Planck Equation}
\newblock (Springer, Berlin)

\bibitem{Bergenholtz02}
Bergenholtz J, Brady J~F  and Vicic M
\newblock  2002 {\em J Fluid Mech} {\bf 456} 239

\bibitem{Derks04}
Derks D, Wisman H , van Blaaderen A and  Imhof A
\newblock 2004 {\em
J Phys Condens Matter} {\bf16} S3917

\bibitem{Petekidis02b}
Petekidis G, Moussaid A  and Pusey P~N
\newblock  2002 {\em Phys Rev E} {\bf 66} 051402

\bibitem{Strating99}
Strating P
\newblock  1999 {\em Phys Rev E} {\bf 59} 2175

\bibitem{Senff99}
Senff H, Richtering W, Norhausen C, Weiss W  and Ballauff M
\newblock  1999 {\em Langmuir} {\bf 15} 102

\bibitem{Fuchs02}
Fuchs M and Cates M~E
\newblock  2002 {\em Phys Rev Lett} {\bf 89} 248304

\bibitem{Goetze91b}
G{\"o}tze W
\newblock  1991 {\em Liquids, Freezing and Glass Transition} eds{} Hansen J-P,
  Levesque D  and Zinn-Justin J
\newblock (North-Holland, Amsterdam) p 287

\bibitem{Cates89}
Cates M~E and Milner S~T
\newblock  1989 {\em Phys Rev Lett} {\bf 62} 1856

\bibitem{Fuchs03}
Fuchs M and Cates M~E
\newblock  2003 {\em Faraday Disc} {\bf 123} 267

\bibitem{Fuchs03b}
Fuchs M and Cates M~E
\newblock  2002 {\em J Phys: Condens Matter} {\bf 15} S401

\bibitem{Kawasaki73b}
Kawasaki K and Gunton J~D
\newblock  1973 {\em Phys Rev A} {\bf 8} 2048

\bibitem{Morriss87}
Morriss G~P and Evans D~J
\newblock  1987 {\em Phys Rev A} {\bf 35} 792

\bibitem{Miyazaki02}
Miyazaki K and Reichman D~R
\newblock  2002 {\em Phys Rev E} {\bf 66} 050501(R)

\bibitem{Miyazaki04}
Miyazaki K, Reichman D~R  and Yamamoto R
\newblock  2004 {\em Phys Rev E} {\bf 70} 011501

\bibitem{Szamel04}
Szamel G
\newblock  2004 {\em Phys Rev Lett} {\bf 93} 178301

\bibitem{McLennan88}
J. A. McLennan J~A
\newblock 1988 {\em Introduction to non-equilibrium statistical mechanics}
\newblock (Prentice Hall, New York)

\bibitem{Forster75}
Forster D
\newblock  1975 {\em Hydrodynamic Fluctuations, Broken Symmetry, and
  Correlation Functions}
\newblock (WA Benjamin, Reading, MA)

\bibitem{Pusey91}
Pusey P~N
\newblock  1991 {\em Liquids, Freezing and Glass Transition} eds{} Hansen J-P,
  Levesque D  and Zinn-Justin J
\newblock (North-Holland, Amsterdam) p 763

\bibitem{Megen93}
van Megen W and Underwood S~M
\newblock  1993 {\em Phys Rev Lett} {\bf 70} 2766

\bibitem{Megen94b}
van Megen W and Underwood S~M
\newblock  1994 {\em Phys Rev E} {\bf 49} 4206

\bibitem{Cichocki87}
Cichocki B and Hess W
\newblock  1987 {\em Physica} {\bf A 141} 475

\bibitem{Kawasaki95}
Kawasaki K
\newblock  1995 {\em Physica A} {\bf 215} 61

\bibitem{Tricomi57}
Tricomi F~G
\newblock  1957 {\em Integral Equations}
\newblock (Interscience Publishers, New York)
\bibitem{Latz02}
Latz A
\newblock  2002 {\em J Stat Phys} {\bf 109} 607

\end{thebibliography}

\end{document}